\documentclass[aps,prd,twocolumn,superscriptaddress,showpacs,nofootinbib]{revtex4-2}
\usepackage{graphicx}
\usepackage{dcolumn}
\usepackage{bm}
\usepackage{hyperref}
\usepackage{amsmath}
\usepackage{comment}
\usepackage[utf8]{inputenc}

\usepackage{caption}
\usepackage{subcaption}

\usepackage[dvipsnames]{xcolor}
 
\usepackage{booktabs, cellspace}
\usepackage{siunitx}
\usepackage{hyperref}

\usepackage{graphicx}
\usepackage{amssymb}

\begin{document}
  \title{A note on  the gravitational wave energy spectrum of parabolic and hyperbolic encounters}

\author{Matthias Gr\"obner}
\affiliation{Department of Physics, ETH Zürich,  Otto-Stern-Weg 1, Zürich, Switzerland} 
\affiliation{Physik-Institut, Universität Zürich, Winterthurerstrasse 190, Zürich, Switzerland} 

\author{Philippe Jetzer}
\affiliation{Physik-Institut, Universität Zürich, Winterthurerstrasse 190, Zürich, Switzerland} 

\author{Maria Haney}
\affiliation{Physik-Institut, Universität Zürich, Winterthurerstrasse 190, Zürich, Switzerland}

\author{Shubhanshu Tiwari}
\affiliation{Physik-Institut, Universität Zürich, Winterthurerstrasse 190, Zürich, Switzerland} 

\author{Wako Ishibashi}
\affiliation{Physik-Institut, Universität Zürich, Winterthurerstrasse 190, Zürich, Switzerland}

\begin{abstract}
The first calculation of the frequency spectrum of gravitational wave mass quadrupole radiation for binaries on hyperbolic orbits was performed in \citet{DeVittori2012}. Some shortcomings of their derivation were pointed out, but there are still inaccuracies and supplements that we believe are worthwhile to communicate. In this note we provide a consistent and straightforward exposition of the frequency spectrum in the case of hyperbolic encounters and explicitly determine the parabolic limit, which was not possible with the previous treatments.
\end{abstract}

\maketitle
    
\section{Introduction}
\label{intro}
The first detection of a gravitational wave signal was accomplished by LIGO in 2015 and resulted from a binary black hole merger \cite{Abbott2016a}. Ten detections of binary black hole mergers and one from a merger of two neutron stars were observed in the first and second observing runs of Advanced LIGO and Advanced
Virgo \cite{Abbott2019}. Relevant for the interpretation of these signals are physical quantities such as the energy spectra for binary systems on Keplerian orbits. In case of binary systems on circular and elliptical orbits the resulting energy spectra have already been well studied \cite{Peters1963}. It is also possible that black holes will not end up in bound systems, but instead produce single scattering events. Such unbound interacting compact binaries can produce gravitational wave burst with emitted frequencies falling in the Advanced LIGO and LISA sensitivity range \cite{Garcia-Bellido2018}. Several studies have estimated the event detection rate of Advanced LIGO and LISA, that can be expected from such encounters, to lie in the order of a few events/year/$\text{Gpc}^3$  \cite{Kocsis2006,OLeary2009,Garcia-Bellido2017}. \\

A first detailed calculation of the energy spectrum for hyperbolic orbits was performed in \cite{DeVittori2012}. Some shortcomings in the Fourier transformations occurred and were subsequently corrected by  \cite{Garcia-Bellido2018}. Nevertheless, throughout the treatments of  \cite{DeVittori2012} and \cite{Garcia-Bellido2018}, a mistaken definition of the Hankel function $H_{\alpha}^{(1)}(z)$ is used. Although this mathematical imprecision does not affect the final energy spectrum formula in \cite{Garcia-Bellido2018}, we believe that in view of the rapid development of gravitational wave astronomy it is important to emphasize the mathematical aspects. Therefore this comment contains an improved and more precise exposition of the material of the paper \cite{DeVittori2012}. As the parabolic limit of the hyperbolic result was not undertaken in neither \cite{DeVittori2012} nor \cite{Garcia-Bellido2018}, we present this derivation also here. The parabolic energy spectrum has been inferred from the elliptic energy spectrum in  \cite{Berry2010}. However, in \cite{Berry2010} the domain of validity of some intervening Bessel functions is not correctly used. Therefore we will in this note also revisit this derivation and highlight thereby the similarities between the elliptic and hyperbolic case.

\section{Energy spectrum for binaries on hyperbolic orbits \label{hyperbolicfreq}}

The energy spectrum of gravitational wave mass quadrupole radiation for binaries on an elliptic
Keplerian orbit was first calculated in \cite{Peters1963}. By an analytic continuation argument the structure of the hyperbolic frequency spectrum can be inferred from the elliptic one. However, it is more instructive to present a detailed outline of the calculation first. We will come back to the algebraic similarities between the two cases when taking the respective parabolic limits. We will use the notations and conventions of the multipole decomposition formalism in linearized general relativity as given by \cite{Thorne1980} (in particular $G=c=1$). \\

The coordinates on an hyperbolic orbit can be parametrized as 
\begin{align}
x &= a(\cosh{u}-e), \\
y &= b \sinh{u} = -a (e^2 -1)^\frac12 \sinh{u},
\end{align} 
where the hyperbolic anomaly $u$ satisfies the hyperbolic Kepler equation
\begin{equation}
e \sinh(u) - u = \Omega t \equiv \frac{\omega}{\nu} t. \label{Keplereq_hyperbolic}
\end{equation} 
Recall that for hyperbolic orbits the semi-major axis $a$ is strictly negative and therefore the angular frequency is taken as $\Omega = \sqrt\frac{M}{-a^3}  = \left( \frac{M (e-1)^3}{r_p^3} \right)^{1/2}$. The second mass moments have accordingly the forms
\begin{align}
\begin{split}
\mathcal{G}_{11} &=  \mu  a^2(e^2 - 2e\cosh{u} + \frac{1}{2} \cosh(2u) + \frac{1}{2} ),\\
\mathcal{G}_{12} = \mathcal{G}_{21} &=  - \mu ab (e\sinh{u} - \frac{1}{2} \sinh(2u)),\\
\mathcal{G}_{22} &=  \mu  \left(\frac{b^2}{2} \cosh(2u) - \frac{b^2}{2} \right). 
\end{split}
\end{align} 

We can decompose the total radiated energy in the frequency domain as follows\footnote{Note that other authors, $e.g.$ \cite{Garcia-Bellido2018}, define $P(\omega)$ through \\ $E \equiv \frac{1}{\pi} \int_{0}^{\infty} d\omega \, P(\omega)$.}

\begin{align}
E &=  \frac{8\pi}{75} \int_{-\infty}^{\infty} dt \, \sum_{m=-2}^{2}  \, \left|  {}^{(3)}\!  \mathcal{G}_{a_1 a_2} \left(\mathcal{Y}^{2m}_{a_1 a_2}\right)^*  \right|^2  \\
&=  \frac{1}{2\pi}\frac{8\pi}{75} \int_{-\infty}^{\infty} d\omega \, \sum_{m=-2}^{2}  \, \left| \widehat{\left[ {}^{(3)}\!  \mathcal{G}_{a_1 a_2} \left(\mathcal{Y}^{2m}_{a_1 a_2}\right)^* \right]} \right|^2 \\
&\equiv \frac{1}{2} \int_{-\infty}^{\infty} d\omega \, P(\omega) =  \int_{0}^{\infty} d\omega \, P(\omega), \label{EandPhyperbolic}
\end{align}
where the second equality follows from Plancherel theorem and the hat represents the Fourier transform.
We carry out the derivative
\begin{align}
P(\omega) = \frac{8}{75} \, \omega^6 \sum_{m=-2}^{2}  \left| \left(\mathcal{Y}^{2m}_{a_1 a_2}\right)^* \widehat{ \mathcal{G}_{a_1 a_2} } \right|^2, \label{P(omega)}
\end{align}

and then it remains to calculate the Fourier transform of $\mathcal{G}_{a_1 a_2}$, that is
\begin{align*}
\widehat{  \mathcal{G}_{a_1 a_2} } &= \int_{-\infty}^\infty du \, \frac{dt(u)}{du}  \, e^{-i\omega t(u)} \, \mathcal{G}_{a_1 a_2}  \\
&= - \frac{1}{i\omega}\int_{-\infty}^\infty du \, \frac{d}{du}\left( e^{-i\omega t(u)} \right)  \, \mathcal{G}_{a_1 a_2}. 
\end{align*}
We will outline the remaining calculation for the $\sinh(nu)$ terms
\begin{align*}
\widehat{ \sinh(nu) } &=  - \frac{1}{i\omega}\int_{-\infty}^\infty du \, \frac{d}{du}\left( e^{-i\omega t(u)} \right)  \, \sinh(nu) \\
&= \frac{n}{i\omega}\int_{-\infty}^\infty du \,  e^{-i\omega t(u)} \, \cosh(nu) \\
&= \frac{n}{2i\omega}  \int_{-\infty}^\infty du \, \left( e^{-i  \nu e\sinh{u} + (i  \nu + n )u} \right. \\
 &\,\,\,\,\,\,\,\,\,\,\,\,\,\,\,\,\,\,\,\,\,\,\,\,\,\,\,\,\,\, \left. + \, e^{-i  \nu e\sinh{u} + (i  \nu -n )u} \right);
\end{align*}
the second equality follows from partial integration and the vanishing of the boundary terms. This result can be expressed in terms of modified Bessel functions of the second kind $K_{\alpha}(x)$. One possible integral representation of $K_{\alpha}(x)$ has the form (see $e.g.$  page 182 in the Bessel function treatise \cite{Watson}) 
\begin{equation}
K_{\alpha}(x) = \frac{1}{2} e^{\frac{1}{2}\alpha\pi i} \int_{-\infty}^\infty dt \, e^{-ix \sinh{t} + \alpha t}. \label{modbesselK_fourier}
\end{equation}
This formula is valid for positive $x$. In \cite{Garcia-Bellido2018} a slightly incorrect definition of the Hankel functions of the first kind is used for rewriting the Fourier integrals. A correct integral representation of the Hankel functions of the first kind is 
\begin{equation}
H^{(1)}_\nu(z) = \frac{1}{\pi i} \int_{\mathcal{C}} dt \, e^{ z \sinh(t) - \nu t}, \label{step}
\end{equation}
which is valid for $|\text{arg}(z)| < \frac{\pi}{2}$ and where the contour $\mathcal{C}$ consists of the three line segments $(-\infty,0] \cup [0,i\pi] \cup [i\pi, i\pi + \infty)$. In \cite{Garcia-Bellido2018} the contour of integration is instead taken as $(-\infty,\infty)$. Only in the limit of purely imaginary argument and with positive $x$, the defining contour of the Hankel functions $H_{\alpha}^{(1)}(ix)$ can be deformed to the one of the modified Bessel functions of the second kind. Fortunately enough, this is precisely the limit needed for the above Fourier integrals, so eventually the final formulae in \cite{Garcia-Bellido2018}  conform with our final formulae. \\*

Using the integral representation (\ref{modbesselK_fourier}) for $K_{\alpha}(x)$ we can write for the Fourier transforms\footnote{Strictly speaking, when $\text{Re}(\alpha) \in \mathbb{Z}$, the function $K_{\alpha}(x)$ has to be interpreted as the limiting value $\lim_{\beta \rightarrow \alpha} K_{\beta}(x)$ for $\text{Re}(\beta) \in \mathbb{R} \setminus \mathbb{Z}$.}
\begin{align*}
\widehat{ \sinh(nu) } &= \frac{n}{i\omega} e^{\frac{1}{2} \nu \pi}  e^{-\frac{1}{2} n\pi i} \left[  K_{i \nu + n } \left( \nu e \right) +e^{i \pi n} \, K_{i \nu - n } \left( \nu e \right) \right].
\end{align*}

A similar calculation gives 
\begin{align*}
\widehat{ \cosh(nu) } &= \frac{n}{i\omega} e^{\frac{1}{2} \nu \pi}  e^{-\frac{1}{2} n\pi i} \left[  K_{i \nu + n } \left( \nu e \right) - e^{i \pi n} \, K_{i \nu - n } \left( \nu e \right) \right].
\end{align*}

By the use of the recurrence relations
\begin{align}
\begin{split}
K_\alpha(x) &= - \frac{x}{2\alpha} [K_{\alpha-1}(x) - K_{\alpha+1}(x)], \\
K'_\alpha(x) &= -\frac{1}{2} [K_{\alpha-1}(x) + K_{\alpha+1}(x)] ,
\end{split}
\end{align}

we can express the Fourier coefficients entirely in terms of $K'_\alpha(x)$ and its derivative. Combining the above we obtain for the relevant terms for $\hat{\mathcal{G}}$
\begin{align}
\begin{split}
\widehat{ \sinh(u) } &= - \frac{2i}{\omega e} \, e^{\frac{1}{2} \nu \pi} \, K_{i \nu } \left( \nu e \right), \\
\widehat{ \sinh(2u) } &=  - \frac{2}{i\omega} e^{\frac{1}{2} \nu \pi}  \left[  \left( \frac{- 4  }{ e^2} + 2 \right) K_{i \nu }\left( \nu e \right) - \frac{4}{\nu e} K'_{i \nu } \left( \nu e \right) \right], \\
\widehat{ \cosh(u) } &=  \frac{2}{\omega} \, e^{\frac{1}{2} \nu \pi}  \, K'_{i \nu } \left( \nu e \right) , \\
\widehat{ \cosh(2u) } &=  \frac{8}{e \omega} \, e^{\frac{1}{2} \nu \pi}   \left[ K'_{i \nu}\left( \nu e \right) - \frac{1}{ \nu e } K_{i \nu}\left( \nu e \right) \right].
\end{split} \label{fouriertransformsK}
\end{align}

By plugging in the explicit expressions of the Fourier transforms (\ref{fouriertransformsK}) into equation (\ref{P(omega)}) for the radiated power in the frequency domain we find 
\begin{align}
P(\omega) &= \frac{8}{15\pi} \, \omega^4 \, \mu^2 \, \left( \frac{r_p}{1-e} \right)^4  e^{\nu \pi} \,  \Pi, \label{hyperbolic_1}
\end{align}

where 
\begin{eqnarray}
\begin{split}
\Pi &=  \left|  \frac{1}{\nu}   K_{i \nu}\left( \nu e \right)  \right|^2  \\
&\,\,\, + 3 \left|  2  K'_{i \nu } \frac{ 
1 - e^2 }{e }  -  \frac{ 2 - e^2 }{ \nu e^2 }   K_{i \nu}\left( \nu e \right)  \right|^2 \\
&\,\,\, + 12  \left|   \frac{(e^2 -1)^\frac32}{e^2}  \, K_{i \nu } \left( \nu e \right)  -  \frac{(e^2 -1)^\frac12}{\nu e  } K'_{i \nu } \left( \nu e \right)  \right|^2 . \label{Pi}
\end{split}
\end{eqnarray}
We obtain after factoring out terms  
\begin{align}
    \begin{split}
   P(\omega) &= \frac{32}{15\pi} \, \frac{\omega^4  \mu^2}{\nu^2} \, \left( \frac{r_p}{e(1-e)} \right)^4  e^{\nu \pi} \\
    &\times \left\lbrace K^2_{i \nu }\left( \nu e \right) \left[ 3 - 3 \nu^2 + 3 e^6 \nu^2 + e^4 (1 - 9 \nu^2) + e^2 (-3 + 9 \nu^2) \right]  \right. \\
    &\,\,\,\,\,\,\,\,\,\,\left. + K'^2_{i \nu }\left( \nu e \right) \left[ 3 e^2 (-1 + e^2) (1 + (-1 + e^2) \nu^2) \right] \right. \\
        &\,\,\,\,\,\,\,\,\,\,\left. + K_{i \nu }\left( \nu e \right)  K'_{i \nu }\left( \nu e \right) \left[  -3 e \nu (4 - 7 e^2 + 3 e^4) \right] \right\rbrace .
    \end{split} \label{Pn_hyperbolic}
\end{align}

The power emitted at zero frequency is non-zero \mbox{for $e>1$;} this effect is called gravitational wave memory. To see this, we expand $K_{0}\left( \nu e \right)$ and  $K'_{0}\left( \nu e \right)$ to first order in $\nu$ around $\nu = 0$. The result is (see $e.g.$ formula (9.6.13) in \cite{Abramowitz}) 
\begin{align*}
K_{0} \left( \nu e \right) &= - \gamma - \log\left(\frac{e}{2}\right) - \log(\nu) + \mathcal{O}(\nu^2), \\
K'_{0} \left( \nu e \right) &=   - \frac{1}{e\nu} + 
   \frac{e\nu}{4} \left(1 - 2 \gamma + 2  \log\left(\frac{2	}{e\nu}\right)  \right) + \mathcal{O}(\nu^2), 
\end{align*}

where $\gamma$ denotes Euler's constant. Inserting the above expressions into the equations (\ref{hyperbolic_1}) - (\ref{Pi}) and then taking the limit towards $\omega = 0$ gives 
\begin{equation}
\lim_{\omega \to 0} P(\omega) = \frac{32}{5\pi} \frac{\mu^2 M^2 (e^2 -1) }{a^2 e^4}. \label{powerzweofreq}
\end{equation}

The gravitational memory effect that results from hyperbolic encounters has been calculated up to 1.5 post-Newtonian accuracy  in \cite{DeVittori2014}. They find that only the cross polarization state of the radiation field contributes to the memory effect and that in the Newtonian limit this state behaves as $|h_\times|^2 \propto \frac{\mu^2 M^2}{a^2} \frac{e^2-1}{e^4}$, which agrees with expression (\ref{powerzweofreq}).  \\*

\section{Energy spectrum for binaries on parabolic orbits \label{parabolicfreq}}In this section we determine the parabolic energy spectrum by taking the appropriate limit of the corresponding elliptic and
hyperbolic result. Both approaches agree. \\*

The total energy emitted into the $n$-harmonic $E(n) = \frac{2 \pi}{\Omega} P(n)$, where $P(n)$ is the power emitted into the $n$-harmonic, during one elliptic orbit is equivalent the energy spectrum $\frac{dE(\omega)}{d\omega}$ multiplied by $\Omega$, 
\begin{equation}
\frac{dE(\omega)}{d\omega} \bigg\rvert_{\text{elliptic}} = \frac{2 \pi}{\Omega^2} P(n),
\end{equation}

and given that the involved expressions are subject to the substitution
\begin{equation}
n \mapsto \frac{\omega}{\Omega} = \frac{r_p^{3/2}}{M^{1/2}} \frac{\omega}{(1-e)^{3/2}} \equiv \tilde{\nu} \frac{1}{(1-e)^{3/2}}. \label{subst}
\end{equation}

We recast the elliptic energy spectrum (for the power spectrum $P(n)$ see $e.g.$ \cite{Peters1963}) to the form 
\begin{align}
\frac{dE(\omega)}{d\omega} \bigg\rvert_{\text{elliptic}} &= \frac{8 \pi}{15} \mu^2 \omega^4   \,   \frac{r_p^4}{(1-e)^4}  \Psi, \label{elliptic1}
\end{align}
where 
\begin{align}
\begin{split}
\Psi &=   \left|   \frac{(1-e)^{3/2}}{ \tilde{\nu}} J_n(ne)   \right|^2  \\
&\,\,\, + 3 \left|   2 J'_n(ne)  \frac{1-e^2}{e}    - \frac{(1-e)^{3/2}}{ \tilde{\nu}}  \frac{2 -e^2 }{e^2} J_n(ne)   \right|^2 \\
&\,\,\, + 12  \left|    \frac{(1-e^2)^\frac32}{e^2}    J_n(ne) - \frac{(1-e^2)^\frac12 (1-e)^{3/2}}{e \tilde{\nu}} J'_n(ne)  \right|^2 .
\end{split}  \label{Psi1}
\end{align}

In the limit of $e\rightarrow 1$, $n \rightarrow \infty$, we need $J_{n}(nz)$ and $J'_{n}(nz)$  as $n \rightarrow \infty$. We have the following asymptotic expansion for $J_{n}(nz)$ (this Bessel function expansion and the ones to follow are derived in \cite{Olver1954})
\begin{align}
J_n(nz) &\sim  \left( \frac{4 \zeta}{1-z^2} \right)^{1/4} \left\{ \frac{\text{Ai}(n^{2/3} \zeta )}{n^{1/3}} \sum_{s=0}^{\infty} \frac{A_s(\zeta)}{n^{2s}}  \right. \notag\\
 & \,\,\,\,\,\,\,\,\,\,\,\,\,\,\,\,+ \left.  \frac{\text{Ai}'(n^{2/3} \zeta )}{n^{5/3}} \sum_{s=0}^{\infty} \frac{B_s(\zeta)}{n^{2s}} \right\},   \label{asymptoticexpansionJ}
\end{align}
as $n \rightarrow \infty$ and provided that $|\text{arg}(z)| < \pi$. In the above formula $\text{Ai}$ denotes the Airy function of the first kind and $\zeta$ is given by 
\begin{eqnarray}
\frac{2}{3} (-\zeta)^{3/2} = (z^2-1)^{1/2} - \arccos\left(\frac{1}{z}\right). \label{zeta}
\end{eqnarray}
The coefficients $A_s$ and $B_s$ are given by
\begin{align*}
A_{s}(\zeta) &= \sum_{m=0}^{2s} b_m \zeta^{-\frac{3}{2}m} U_{2s-m}, \\
 \zeta^\frac12 B_{s}(\zeta) &= - \sum_{m=0}^{2s+1} a_m \zeta^{-\frac{3}{2}m} U_{2s-m+1},
\end{align*}

in which $U_0 = 1$ and with $u=(1-z^2)^{-\frac{1}{2}}$ 
\begin{align*}
U_{s+1} = \frac{1}{2} u^2 (1-u^2) \frac{dU_s}{du} + \frac{1}{8} \int_{0}^u du \, (1-5u^2)  U_s .
\end{align*}
The remaining coefficients are recursively defined by $a_0 = b_0 =1$ and 
\begin{align*}
a_s &= \frac{(2s+1)(2s+3)\cdots(6s-1)}{s! \, (144)^s}, \\
 b_s &= -  \frac{6s +1}{6s-1} a_s.
\end{align*}
\\

Under the condition that $nz < n$, one can find the following asymptotic expansion of $J_n(nz)$ (see \textit{e.g.} page 249 in \cite{Watson})
 \begin{align}
J_n(nz) &\sim \frac{1}{\pi} \left( \frac{2(1-z)}{3z} \right)^\frac12 K_{\frac13}\left( \frac{2^\frac32}{3 z^\frac12} n (1-z)^{\frac32} \right). \label{J_n_limitapprox_stat}
\end{align}
It is easy to verify that in the limit of $z\rightarrow 1$, the leading order term of the expansion (\ref{asymptoticexpansionJ}) coincides with (\ref{J_n_limitapprox_stat}). The disadvantage of formula (\ref{J_n_limitapprox_stat}) is that it is not possible to determine the exact domain of validity. In \cite{Berry2010} formula (\ref{J_n_limitapprox_stat}) is used. However, in \cite{Berry2010} it is also applied to the case when $nz > n$, where the formula strictly speaking does not hold. All asymptotic formulae we use here are more general and have no restrictions on the value of $z$. In addition, as we will see, the general asymptotic formulae make it very transparent why the parabolic limit of both the elliptic and hyperbolic cases coincide.  The Bessel function derivative $J'_n(ne)$ has the asymptotic expansion 
\begin{align}
J'_n(nz) &\sim  - \frac{2}{z} \left( \frac{1-z^2}{4 \zeta} \right)^{1/4} \left\{ \frac{\text{Ai}(n^{2/3} \zeta )}{n^{4/3}} \sum_{s=0}^{\infty} \frac{C_s(\zeta)}{n^{2s}}  \right. \notag\\
 & \,\,\,\,\,\,\,\,\,\,\,\,\,\,\,\,+ \left.  \frac{\text{Ai}'(n^{2/3} \zeta )}{n^{2/3}} \sum_{s=0}^{\infty} \frac{D_s(\zeta)}{n^{2s}} \right\},   \label{asymptoticexpansionJ'}
\end{align}

where 
\begin{align*}
C_s(\zeta) &= \chi(\zeta) A_s(\zeta) + A'_s(\zeta) + \zeta B_s(\zeta) , \\
D_s(\zeta) &=  A_s(\zeta) + \chi(\zeta) B_{s-1}(\zeta) +  B'_{s-1}(\zeta),  
\end{align*}
and 
\begin{align*}
\chi(\zeta) = \frac{4 - z^2  \left( \frac{4 \zeta}{1-z^2} \right)^{3/2} }{16\zeta}.
\end{align*}
Before giving the final expression for the parabolic case as derived as a limit from the elliptic
case, we consider the corresponding limit from the
hyperbolic energy spectrum. To obtain the parabolic energy spectrum, we need to perform in the expressions (\ref{hyperbolic_1})  - (\ref{Pi}) the limit as \mbox{$\nu  = \frac{\tilde{\nu}}{(e-1)^{3/2}} \rightarrow \infty$.} The desired asymptotic expansions of the modified Bessel functions of the second kind are 
\begin{align}
K_{i\nu}(\nu e) &\sim   \pi  e^{-\frac{\nu \pi}{2}}   \left( \frac{4 \zeta}{1-e^2} \right)^{1/4} \notag \\ &\times \left\{ \frac{\text{Ai}(- \nu^{2/3} \zeta )}{\nu^{1/3}} \sum_{s=0}^{\infty} (-1)^{s} \frac{A_s(\zeta)}{\nu^{2s}}  \right. \notag\\
  & \,\,\,\,\,\,\,\,\,\,\,\,\,\,\,\,+ \left.  \frac{\text{Ai}'(-\nu^{2/3} \zeta )}{\nu^{5/3}} \sum_{s=0}^{\infty} (-1)^{s}  \frac{B_s(\zeta)}{\nu^{2s}} \right\}, \label{asymptoticexpansionK} \\ 
{K'_{i\nu}}(\nu e) &\sim   - \pi  e^{-\frac{\nu \pi}{2}} \frac{2 }{e} \left( \frac{1-e^2}{4 \zeta} \right)^{1/4} \notag \\
&\times \left\{ \frac{\text{Ai}(- \nu^{2/3} \zeta )}{\nu^{4/3}} \sum_{s=0}^{\infty} (-1)^{s}  \frac{C_s(\zeta)}{\nu^{2s}} \right. \notag\\
  & \,\,\,\,\,\,\,\,\,\,\,\,\,\,\,\,- \left. \frac{\text{Ai}'(-\nu^{2/3} \zeta )}{\nu^{2/3}} \sum_{s=0}^{\infty} (-1)^{s}  \frac{D_s(\zeta)}{\nu^{2s}} \right\}. \label{asymptoticexpansionK'}
\end{align}

We will now specialize the above expressions to the limit of $e \rightarrow 1$. A power counting argument shows that, in this limit, only the leading term in each Bessel function expansion evaluates to a non-zero value in the elliptic spectrum (\ref{elliptic1}) - (\ref{Psi1}) and in the hyperbolic \mbox{spectrum (\ref{hyperbolic_1})  - (\ref{Pi}).} Expanding equation (\ref{zeta}) around $z=1$ yields the following value for $\zeta$ 
\begin{equation}
\zeta = 2^{\frac13} (1-z).
\end{equation}

Applying the above limit of $\zeta$ and restoring $n = \tilde{\nu} \frac{1}{(1-e)^{3/2}}$  and $\nu  = \frac{\tilde{\nu}}{(e-1)^{3/2}}$  in the respective variables we find the following expressions for the leading terms of the Bessel function expansions (\ref{asymptoticexpansionJ}), (\ref{asymptoticexpansionJ'}), (\ref{asymptoticexpansionK}) and (\ref{asymptoticexpansionK'}) 
\begin{align}
J_n(ne) &\sim \left( \frac{2}{\tilde{\nu}} \right)^{1/3}  (1-e)^\frac12 \, \text{Ai}( 2^{\frac13} \tilde{\nu}^{\frac23}  ), \label{J_n_limit1} \\
J'_n(ne) &\sim -  \left( \frac{2}{\tilde{\nu}} \right)^{2/3}    \frac{1-e}{e} \, \text{Ai}'( 2^{\frac13} \tilde{\nu}^{\frac23}  ), \label{J'_n_limit1} \\
K_{i\nu}(\nu e) &\sim \pi e^{-\frac{\nu \pi}{2}} \left( \frac{2}{\tilde{\nu}} \right)^{1/3}  (e-1)^\frac12 \, \text{Ai}(2^{\frac13} \tilde{\nu}^{\frac23}   ), \label{K_limit1} \\
K'_{i\nu}(\nu e) &\sim  \pi e^{-\frac{\nu \pi}{2}} \left( \frac{2}{\tilde{\nu}} \right)^{2/3}    \frac{e-1}{e} \, \text{Ai}'( 2^{\frac13} \tilde{\nu}^{\frac23}  ). \label{K'_n_limit1}
\end{align}

An inspection of the structure of the energy spectra  \mbox{(\ref{elliptic1}) - (\ref{Psi1})} and (\ref{hyperbolic_1})  - (\ref{Pi}) and of the above expansions makes it evident that the parabolic limit of the elliptic energy spectrum and the parabolic limit of the hyperbolic energy spectrum coincide. Indeed, both limits give the parabolic energy spectrum as 
\begin{align}
\begin{split}
\frac{dE(\omega)}{d\omega} \bigg\rvert_{\text{parabolic}} &= \frac{128 \pi}{5 } \, \omega^4 \, \mu^2 \, r_p^4  \, \\ 
 &\times \left\{   \text{Ai}^2( 2^{\frac13} \tilde{\nu}^{\frac23}  ) \left(\frac{2}{\tilde{\nu}} \right)^{2/3} \left( 2 + \frac{1}{12 \tilde{\nu}^2} \right) \right. \\
&\,\,\,\,\,\, \left. +  \, {\text{Ai}'}^2( 2^{\frac13}  \tilde{\nu}^{\frac23}  )  \left(\frac{2}{\tilde{\nu}} \right)^{4/3}  \right. \\
&\,\,\,\,\,\, \left. +  \, \text{Ai}( 2^{\frac13} \tilde{\nu}^{\frac23}  ) \,  {\text{Ai}'}( 2^{\frac13}  \tilde{\nu}^{\frac23}  )  \frac{1}{\tilde{\nu}^2}   \right\} .
\end{split}
\end{align}

Acknowledgement:   MH acknowledges support from Swiss National Science Foundation (SNSF) grant Nr. IZCOZ0-177057. ST is supported by Forschungskredit Nr. FK-19-114. PJ and WI acknowledge support from the University of Zurich.

\bibliographystyle{apsrev}
\bibliography{references}

\begin{thebibliography}{14}
\expandafter\ifx\csname natexlab\endcsname\relax\def\natexlab#1{#1}\fi
\expandafter\ifx\csname bibnamefont\endcsname\relax
  \def\bibnamefont#1{#1}\fi
\expandafter\ifx\csname bibfnamefont\endcsname\relax
  \def\bibfnamefont#1{#1}\fi
\expandafter\ifx\csname citenamefont\endcsname\relax
  \def\citenamefont#1{#1}\fi
\expandafter\ifx\csname url\endcsname\relax
  \def\url#1{\texttt{#1}}\fi
\expandafter\ifx\csname urlprefix\endcsname\relax\def\urlprefix{URL }\fi
\providecommand{\bibinfo}[2]{#2}
\providecommand{\eprint}[2][]{\url{#2}}

\bibitem[{\citenamefont{De~Vittori et~al.}(2012)\citenamefont{De~Vittori,
  Jetzer, and Klein}}]{DeVittori2012}
\bibinfo{author}{\bibfnamefont{L.}~\bibnamefont{De~Vittori}},
  \bibinfo{author}{\bibfnamefont{P.}~\bibnamefont{Jetzer}}, \bibnamefont{and}
  \bibinfo{author}{\bibfnamefont{A.}~\bibnamefont{Klein}},
  \bibinfo{journal}{Phys. Rev. D} \textbf{\bibinfo{volume}{86}},
  \bibinfo{pages}{044017} (\bibinfo{year}{2012}).

\bibitem[{\citenamefont{Abbott et~al.}(2016)\citenamefont{Abbott, Abbott,
  Abbott, Abernathy, Acernese, Ackley, Adams, Adams, Addesso, Adhikari
  et~al.}}]{Abbott2016a}
\bibinfo{author}{\bibfnamefont{B.~P.} \bibnamefont{Abbott}},
  \bibinfo{author}{\bibfnamefont{R.}~\bibnamefont{Abbott}},
  \bibinfo{author}{\bibfnamefont{T.~D.} \bibnamefont{Abbott}},
  \bibinfo{author}{\bibfnamefont{M.~R.} \bibnamefont{Abernathy}},
  \bibinfo{author}{\bibfnamefont{F.}~\bibnamefont{Acernese}},
  \bibinfo{author}{\bibfnamefont{K.}~\bibnamefont{Ackley}},
  \bibinfo{author}{\bibfnamefont{C.}~\bibnamefont{Adams}},
  \bibinfo{author}{\bibfnamefont{T.}~\bibnamefont{Adams}},
  \bibinfo{author}{\bibfnamefont{P.}~\bibnamefont{Addesso}},
  \bibinfo{author}{\bibfnamefont{R.~X.} \bibnamefont{Adhikari}},
  \bibnamefont{et~al.} (\bibinfo{collaboration}{LIGO Scientific Collaboration
  and Virgo Collaboration}), \bibinfo{journal}{Phys. Rev. Lett.}
  \textbf{\bibinfo{volume}{116}}, \bibinfo{pages}{061102}
  (\bibinfo{year}{2016}),
  \urlprefix\url{https://link.aps.org/doi/10.1103/PhysRevLett.116.061102}.

\bibitem[{\citenamefont{Abbott et~al.}(2019)\citenamefont{Abbott, Abbott,
  Abbott, Abraham, Acernese, Ackley, Adams, Adhikari, Adya, Affeldt
  et~al.}}]{Abbott2019}
\bibinfo{author}{\bibfnamefont{B.~P.} \bibnamefont{Abbott}},
  \bibinfo{author}{\bibfnamefont{R.}~\bibnamefont{Abbott}},
  \bibinfo{author}{\bibfnamefont{T.~D.} \bibnamefont{Abbott}},
  \bibinfo{author}{\bibfnamefont{S.}~\bibnamefont{Abraham}},
  \bibinfo{author}{\bibfnamefont{F.}~\bibnamefont{Acernese}},
  \bibinfo{author}{\bibfnamefont{K.}~\bibnamefont{Ackley}},
  \bibinfo{author}{\bibfnamefont{C.}~\bibnamefont{Adams}},
  \bibinfo{author}{\bibfnamefont{R.~X.} \bibnamefont{Adhikari}},
  \bibinfo{author}{\bibfnamefont{V.~B.} \bibnamefont{Adya}},
  \bibinfo{author}{\bibfnamefont{C.}~\bibnamefont{Affeldt}},
  \bibnamefont{et~al.} (\bibinfo{collaboration}{LIGO Scientific Collaboration
  and Virgo Collaboration}), \bibinfo{journal}{Phys. Rev. X}
  \textbf{\bibinfo{volume}{9}}, \bibinfo{pages}{031040} (\bibinfo{year}{2019}),
  \urlprefix\url{https://link.aps.org/doi/10.1103/PhysRevX.9.031040}.

\bibitem[{\citenamefont{Peters and Mathews}(1963)}]{Peters1963}
\bibinfo{author}{\bibfnamefont{P.~C.} \bibnamefont{Peters}} \bibnamefont{and}
  \bibinfo{author}{\bibfnamefont{J.}~\bibnamefont{Mathews}},
  \bibinfo{journal}{Phys. Rev.} \textbf{\bibinfo{volume}{131}},
  \bibinfo{pages}{435} (\bibinfo{year}{1963}).

\bibitem[{\citenamefont{García-Bellido and
  Nesseris}(2018)}]{Garcia-Bellido2018}
\bibinfo{author}{\bibfnamefont{J.}~\bibnamefont{García-Bellido}}
  \bibnamefont{and} \bibinfo{author}{\bibfnamefont{S.}~\bibnamefont{Nesseris}},
  \bibinfo{journal}{Physics of the Dark Universe}
  \textbf{\bibinfo{volume}{21}}, \bibinfo{pages}{61 } (\bibinfo{year}{2018}),
  ISSN \bibinfo{issn}{2212-6864}.

\bibitem[{\citenamefont{Kocsis et~al.}(2006)\citenamefont{Kocsis, Gaspar, and
  Marka}}]{Kocsis2006}
\bibinfo{author}{\bibfnamefont{B.}~\bibnamefont{Kocsis}},
  \bibinfo{author}{\bibfnamefont{M.~E.} \bibnamefont{Gaspar}},
  \bibnamefont{and} \bibinfo{author}{\bibfnamefont{S.}~\bibnamefont{Marka}},
  \bibinfo{journal}{The Astrophysical Journal} \textbf{\bibinfo{volume}{648}},
  \bibinfo{pages}{411} (\bibinfo{year}{2006}),
  \urlprefix\url{https://doi.org/10.1086%2F505641}.

\bibitem[{\citenamefont{O'Leary et~al.}(2009)\citenamefont{O'Leary, Kocsis, and
  Loeb}}]{OLeary2009}
\bibinfo{author}{\bibfnamefont{R.~M.} \bibnamefont{O'Leary}},
  \bibinfo{author}{\bibfnamefont{B.}~\bibnamefont{Kocsis}}, \bibnamefont{and}
  \bibinfo{author}{\bibfnamefont{A.}~\bibnamefont{Loeb}},
  \bibinfo{journal}{Monthly Notices of the Royal Astronomical Society}
  \textbf{\bibinfo{volume}{395}}, \bibinfo{pages}{2127} (\bibinfo{year}{2009}),
  ISSN \bibinfo{issn}{0035-8711},
  \eprint{http://oup.prod.sis.lan/mnras/article-pdf/395/4/2127/2931749/mnras0395-2127.pdf},
  \urlprefix\url{https://doi.org/10.1111/j.1365-2966.2009.14653.x}.

\bibitem[{\citenamefont{García-Bellido and
  Nesseris}(2017)}]{Garcia-Bellido2017}
\bibinfo{author}{\bibfnamefont{J.}~\bibnamefont{García-Bellido}}
  \bibnamefont{and} \bibinfo{author}{\bibfnamefont{S.}~\bibnamefont{Nesseris}},
  \bibinfo{journal}{Physics of the Dark Universe}
  \textbf{\bibinfo{volume}{18}}, \bibinfo{pages}{123 } (\bibinfo{year}{2017}),
  ISSN \bibinfo{issn}{2212-6864},
  \urlprefix\url{http://www.sciencedirect.com/science/article/pii/S2212686417300638}.

\bibitem[{\citenamefont{Berry and Gair}(2010)}]{Berry2010}
\bibinfo{author}{\bibfnamefont{C.~P.~L.} \bibnamefont{Berry}} \bibnamefont{and}
  \bibinfo{author}{\bibfnamefont{J.~R.} \bibnamefont{Gair}},
  \bibinfo{journal}{Phys. Rev. D} \textbf{\bibinfo{volume}{82}},
  \bibinfo{pages}{107501} (\bibinfo{year}{2010}),
  \urlprefix\url{https://link.aps.org/doi/10.1103/PhysRevD.82.107501}.

\bibitem[{\citenamefont{Thorne}(1980)}]{Thorne1980}
\bibinfo{author}{\bibfnamefont{K.~S.} \bibnamefont{Thorne}},
  \bibinfo{journal}{Rev. Mod. Phys.} \textbf{\bibinfo{volume}{52}},
  \bibinfo{pages}{299} (\bibinfo{year}{1980}),
  \urlprefix\url{https://link.aps.org/doi/10.1103/RevModPhys.52.299}.

\bibitem[{\citenamefont{Watson}(1966)}]{Watson}
\bibinfo{author}{\bibfnamefont{G.~N.} \bibnamefont{Watson}},
  \emph{\bibinfo{title}{A Treatise on the Theory of Bessel Functions}}
  (\bibinfo{publisher}{Cambridge University Press}, \bibinfo{year}{1966}).

\bibitem[{\citenamefont{Abramowitz and Stegun}(1965)}]{Abramowitz}
\bibinfo{author}{\bibfnamefont{M.}~\bibnamefont{Abramowitz}} \bibnamefont{and}
  \bibinfo{author}{\bibfnamefont{I.}~\bibnamefont{Stegun}},
  \emph{\bibinfo{title}{Handbook of Mathematical Functions}}
  (\bibinfo{publisher}{DOVER PUBN INC}, \bibinfo{year}{1965}).

\bibitem[{\citenamefont{De~Vittori et~al.}(2014)\citenamefont{De~Vittori,
  Gopakumar, Gupta, and Jetzer}}]{DeVittori2014}
\bibinfo{author}{\bibfnamefont{L.}~\bibnamefont{De~Vittori}},
  \bibinfo{author}{\bibfnamefont{A.}~\bibnamefont{Gopakumar}},
  \bibinfo{author}{\bibfnamefont{A.}~\bibnamefont{Gupta}}, \bibnamefont{and}
  \bibinfo{author}{\bibfnamefont{P.}~\bibnamefont{Jetzer}},
  \bibinfo{journal}{Phys. Rev. D} \textbf{\bibinfo{volume}{90}},
  \bibinfo{pages}{124066} (\bibinfo{year}{2014}),
  \urlprefix\url{https://link.aps.org/doi/10.1103/PhysRevD.90.124066}.

\bibitem[{\citenamefont{Olver}(1954)}]{Olver1954}
\bibinfo{author}{\bibfnamefont{F.~W.~J.} \bibnamefont{Olver}},
  \bibinfo{journal}{Philosophical Transactions of the Royal Society of London.
  Series A, Mathematical and Physical Sciences} \textbf{\bibinfo{volume}{247}},
  \bibinfo{pages}{328} (\bibinfo{year}{1954}), ISSN \bibinfo{issn}{00804614},
  \urlprefix\url{http://www.jstor.org/stable/91484}.

\end{thebibliography}

\end{document}